\documentclass{ws-ijmpa}
\newcommand{\be}{\begin{equation}}
\newcommand{\ee}{\end{equation}}
\newcommand{\bea}{\begin{eqnarray}}
\newcommand{\eea}{\end{eqnarray}}
\newcommand{\bean}{\begin{eqnarray*}}
\newcommand{\eean}{\end{eqnarray*}}

\begin{document}

\markboth{F.Q.Wu}{Coupled channel effects
 in $\pi\pi$ S-wave interaction}

\catchline{}{}{}{}{}

\title{Coupled channel effects
 in $\pi\pi$ S-wave interaction}

\author{F.Q.Wu and B.S.Zou}
\address{CCAST (World Laboratory), P.O.Box 8730, Beijing 100080, China ; \\
Institute of High Energy Physics, CAS, Beijing 100049, China}


\maketitle


\begin{abstract}
We study coupled channel effects upon isospin $I=2$ and $I=0$
$\pi\pi$ S-wave interaction. With introduction of the
$\pi\pi\to\rho\rho\to\pi\pi$ coupled channel box diagram
contribution into $\pi\pi$ amplitude in addition to $\rho$ and
$f_2 (1270)$ exchange, we reproduce the $\pi \pi$ I=2 S-wave and
D-wave scattering phase shifts and inelasticities up to 2 GeV
quite well in a K-matrix formalism. For $I=0$ case, the same
$\pi\pi\to\rho\rho\to\pi\pi$ box diagram is found to give the
largest contribution for the inelasticity among all possible
coupled channels including $\pi\pi\to\omega\omega\to\pi\pi$,
$\pi\pi\to K \overline{K}\to\pi\pi$. We also show why the broad
$\sigma$ appears narrower in production processes than in $\pi\pi$
scattering process.
\end{abstract}



\section{INTRODUCTION}

As well known isospin $I=0 \ \ \pi \pi$ S-wave interaction gives a
good place to study the $I=0 \ \ J^{pc}=0^{++}$ particles such as
$\sigma$ and glueball.  However, to really understand the
isoscalar $\pi\pi$ S-wave interaction, one must first understand
the isospin I=2 $\pi\pi$ S-wave interaction due to the following
two reasons:  (1) There are no known s-channel resonances and less
coupled channels in I=2 $\pi\pi$ system, so it is much simpler
than the I=0 $\pi\pi$ S-wave interaction; (2) To extract I=0
$\pi\pi$ S-wave phase shifts from experimental data on
$\pi^+\pi^-\to\pi^+\pi^-$ and $\pi^+\pi^-\to\pi^0\pi^0$ obtained
by $\pi N\to\pi\pi N$ reactions, one needs an input of the I=2
$\pi\pi$ S-wave interaction.

Up to now, experimental information on the I=2 $\pi\pi$ scattering
mainly came from $\pi^+p\to\pi^+\pi^+n$ \cite{Hoogland77} and
$\pi^-d\to\pi^-\pi^-pp$ \cite{Durusoy73} reactions. In previous
analyses, the feature of inelasticity $\eta_0^2$ starting to
deviate from 1 for energies above 1.1 GeV was often overlooked.
Recently, with a K-matrix formalism \cite{lilong}, we show
\cite{Wu} that the feature can be well reproduced by
$\pi\pi$-$\rho\rho$ coupled-channel effect.  Here we extend the
study of the coupled channel effects to the I=0 case which allows
much more coupled channels. We also show why the broad $\sigma$
appears narrower in production processes than in $\pi\pi$
scattering process.

\section{Coupled channel effects in $\pi\pi$ scattering }

\begin{figure}[htbp]
\begin{center}
\includegraphics[width=10cm,height=2cm]{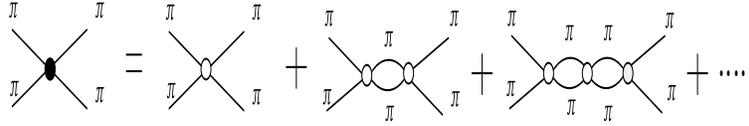}
\end{center}
\caption{Diagrammatic expression for $\pi\pi$ scattering in
K-matrix formalism}
 \label{krk}
\end{figure}

We follow the $K$-matrix formalism as in Ref.\cite{lilong}. For
$\pi\pi$ scattering, the scattering amplitude can be given in
K-matrix formalism as
\be
T_{el}=\frac{K}{1-i\rho K} = K +
K~i\rho~K + K~i\rho~K~i\rho~K + \cdots
\ee
which can be expressed
diagrammatically as in Fig.\ref{krk} for $\pi\pi$ scattering at
low energies with only one opening channel.

\begin{figure}[htbp]
\begin{center}
\includegraphics[width=10cm,height=2cm]{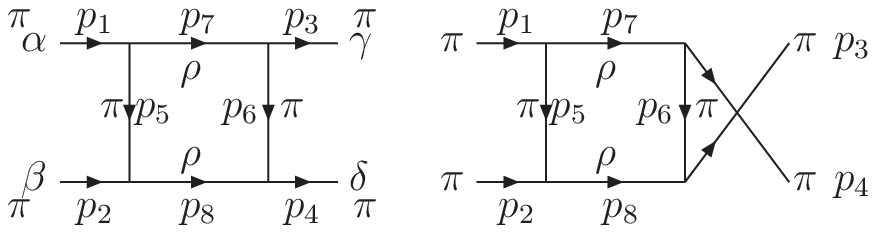}
\end{center}
\caption{The $\pi\pi\to\rho\rho\to\pi\pi$ box diagrams}
 \label{block1}
\end{figure}

For the two-channel case, the two-dimensional $K$ matrix and phase
space $\rho(s)$ matrix are \be K=\left (
\begin{array}{ccc}
K_{11}  &  K_{12}  \\
K_{12}  &  K_{22}
\end{array} \right ),\hspace{1cm}
\rho(s)=\left (
\begin{array}{ccc}
\rho_1(s)  &  0  \\
0  &  \rho_2(s)
\end{array} \right ),
\ee with i=1,2 representing $\pi\pi$ and $\rho\rho$ channel,
respectively.  Ignoring the interaction between $\rho\rho$, we
have $K_{22}=0$; then
\be T_{11}=\frac{K_{11}+i K_{12} \rho_2 K_{21}}{1-i
\rho_1(K_{11}+iK_{12} \rho_2 K_{21} )}, \ee
where $iK_{12} \rho_2 K_{21}$ corresponds to the $\rho\rho$
on-shell contribution\cite{Lu} of the $\pi\pi\to\rho\rho\to\pi\pi$
box diagram as shown in Fig.\ref{block1}.

\begin{figure}
\begin{center}
\includegraphics[scale=0.5]{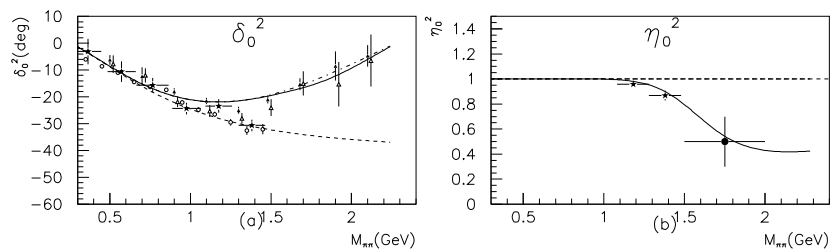}
\end{center}
\vspace{-4cm}
 \caption{The $I=2$ $\pi\pi$ $S$-wave ($\delta_0^2$,
$\eta_0^2$)
 phase shifts and inelasticities. Data are from
Ref.$^{1,2,3}$.  The solid curves represent the total contribution
of $\rho$, $f_2$ exchange and  the box diagram; dot-dashed curves
from $\rho $ and $f_2$ exchange; dashed curves from t-channel
$\rho$ exchange only.}
 \label{phase}
\end{figure}

With $K_{11}$ including contribution from $t$-channel $\rho$ and
$f_2(1270)$ exchange, we found that the basic features of I=2
$\pi\pi$ S-wave phase shifts and inelasticities are well
reproduced as shown in Fig.\ref{phase}. For details, see
Ref.\cite{Wu}.

\begin{figure}[htbp]
\begin{center}
\includegraphics[width=10cm,height=2cm]{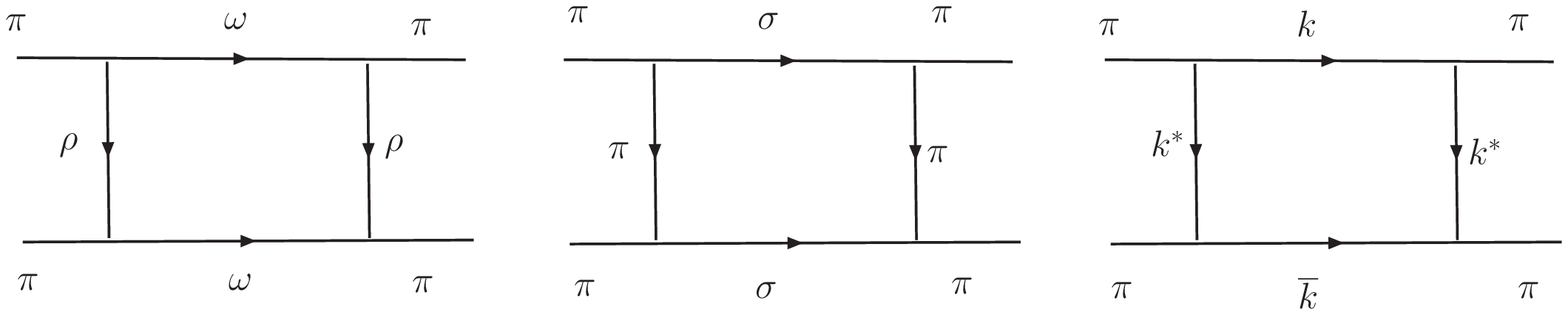}
\end{center}
 \caption{The $\pi\pi\to\omega\omega\to\pi\pi$,
$\pi\pi\to\sigma\sigma\to\pi\pi$ and $\pi\pi\to K
\overline{K}\to\pi\pi$  box diagrams for I=0 case.}
 \label{block2}
\end{figure}

With the success in reproducing the $I=2$ $\pi\pi$ S-wave
scattering, we extend our study of the coupled channel effects to
the isospin I=0 $\pi\pi$ scattering.  Here in additional to the
box diagrams shown in Fig. \ref{block1}, there are more other
coupled channels such as $\pi\pi\to\omega\omega\to\pi\pi$,
$\pi\pi\to\sigma\sigma\to\pi\pi$ and $\pi\pi\to K
\overline{K}\to\pi\pi$ as shown in Fig.\ref{block2}.

\begin{figure}[htbp]
\begin{center}
\includegraphics[scale=0.3]{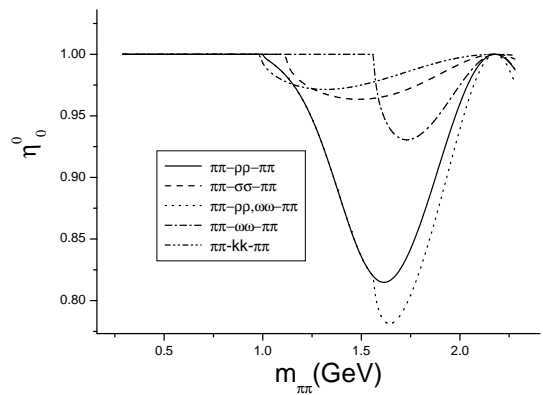}
\end{center}
\vspace{-3cm}
 \caption{The $I=0$ $\pi \pi$ S-wave inelasticity $\eta_{0}^{0}$
without including s-channel resonances in the $I=0$ amplitude.
``$\pi\pi-\rho\rho - \pi \pi$" means including
$\pi\pi-\rho\rho-\pi\pi$ box diagram; ``$\pi\pi-\rho\rho, \omega
\omega- \pi \pi$" means using three-dimensions K matrix to couple
$\pi\pi,\rho\rho$ and $\omega \omega$ channels together. }
 \label{sampl}
\end{figure}

To demonstrate the significance of coupled channel effects in the
I=0 $\pi\pi$ scattering, we  do not include any s-channel
resonances in the $I=0$ amplitude, just introduce $\pi\pi\to
\rho\rho \to\pi\pi$ ,$\pi\pi \to \omega\omega \to \pi \pi$,
$\pi\pi\to K \overline{K}\to\pi\pi$ and $\pi\pi\to \sigma \sigma
\to \pi\pi$ box diagrams respectively into the $I=0$ amplitude
which includes t-channel $\rho$ and $f_{2}(1270)$ exchange
\cite{lilong}. The result shows $\pi\pi\to \rho\rho \to\pi\pi$
(solid line in Fig. \ref{sampl}) gives the most important
contribution. We also use three-dimensions K matrix to couple
$\pi\pi,\rho\rho$, and $\omega \omega$ channels together. The
result is shown by the dotted line in Fig.\ref{sampl}.

\section{$\pi\pi$ S-wave interaction in production processes}

\begin{figure}[htbp]
\begin{center}
\includegraphics[width=10cm,height=2cm]{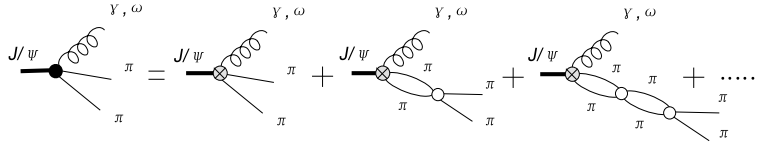}
\end{center}
\caption{Diagrammatic expression for $\pi\pi$ scattering in
K-matrix formalism}
 \label{prk}
\end{figure}

$\pi\pi$ production processes can be regarded as a special case
for coupled channels. For the $\pi\pi$ S-wave interaction in
production processes, it can be expressed diagrammatically as in
Fig.\ref{prk}.

Compared with elastic scattering shown in Fig.\ref{krk}, the only
difference is the first interaction vertex. So the production
amplitude can be expressed similar to Eq.(1) as

\be T_{prod}=\frac{P}{1-i\rho K} = P + P~i\rho~K +
P~i\rho~K~i\rho~K + \cdots \ee

\begin{figure}[htbp]
\begin{center}
\includegraphics[width=8cm,height=6cm]{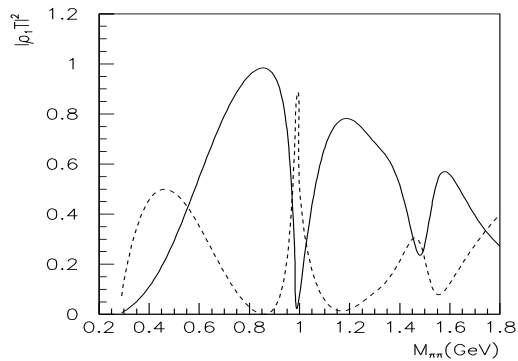}
\end{center}
\caption{Amplitude squared for $\pi\pi$ S-wave in production
process (dashed curve) compared with in elastic scattering (solid
curve), assuming P=1 and K from Ref.$^{9}$. } \label{sigma}
\end{figure}

For the I=0 $\pi\pi$ S-wave scattering at low energies ($\leq
2m_K$, the K-matrix and its corresponding elastic scattering
amplitude $T_{el}$ can be well determined by the $\pi\pi$
scattering phase shift data \cite{CM,Becker,Kaminski}. The solid
line in Fig.\ref{sigma} shows a solution \cite{BSZ} for
$|\rho_1T_{el}|^2$ from fitting the well-known CERN-Munich
$\pi\pi$ scattering data \cite{CM,Becker}. The solution has a
T-matrix pole at $571-i420$ MeV for the broad $\sigma$ which
provides the broad background for two narrow dips caused by its
interference with the narrow $f_0(980)$ and $f_0(1500)$. With the
same K-matrix, if we assume P-matrix to be 1 for $\pi\pi$
production, we can get $|\rho_1T_{prod}|^2$ as shown by the dashed
line in Fig.\ref{sigma}. A much narrower peak at lower energy is
appearing although in fact it has the same broad pole as in the
$\pi\pi$ elastic scattering process. This gives a clear
demonstration why the broad $\sigma$ appears narrower in
production processes than in $\pi\pi$ scattering process. The
reason is $T_{prod}\sim T_{el}/K$ here. This has also been noted
by Ref.\cite{Oset} in a slightly different language.

Some recent analyses \cite{E791,BES} of various $\pi\pi$
production processes gave a much narrower $\sigma$ pole than
Ref.\cite{BSZ,zou1} from $\pi\pi$ scattering. A reason is that it
is assumed a direct production of $\sigma$ with production vertex
$P=1/(m^2_\sigma-m^2_{\pi\pi})$ instead of considering the
$\sigma$ due to final state $\pi\pi$ scattering with $P=1$ or some
smooth function of $m_{\pi\pi}$. With the same production data but
with different production vertex $P$, one will get different
$\sigma$ pole. See Refs.\cite{E791,FOCUS} for an example.



\section*{Acknowledgements}

The work is partly supported by CAS Knowledge Innovation Project
(KJCX2-SW-N02) and the National Natural Science Foundation of
China.


\end{document}